\documentclass{jaa}
\usepackage{graphicx}

\begin{document}

%%paper title
%%For line breaks \\ can be used within title 
\title{Flux-Vortex Pinning and Neutron Star Evolution}

%%author names are separated by comma (,) 
%%use \and before the last author name 
%%use a * along with the number separated by comma
%% for the  author for correspondence
%%\textsuperscript{number} is used for affiliation
%%\affilOne, \affilTwo etc., upto \affilTwentyfive is possible
%%Please note the first letter after \affil is capitalised in the command
%%

\author{M. Ali Alpar\textsuperscript{1}}
\affilOne{\textsuperscript{1}Faculty of Engineering and Natural Sciences, Sabanc{\i} University, 34956 Istanbul, Turkey.}
%\affilTwo{\textsuperscript{2}Department of Q, University Z, Place Pincode, Country.}

%%escape two column mode for title, affiliation and abstract
%%by giving \twocolumn command as shown

\twocolumn[{

\maketitle

%%include \corres to print the corresponding author Email id
\corres{alpar@sabanciuniv.edu}

%%include \msinfo for
%%manuscript information such as
%%received, revised and accepted dates
%%
%%\msinfo{1 January 2015}{1 January 2015}{1 January 2015}

%%abstract
\begin{abstract}
G. Srinivasan et al proposed a simple and elegant explanation for the reduction of the neutron star magnetic dipole moment during binary evolution leading to low mass X-ray binaries and eventually to millisecond pulsars:  Quantized vortex lines in the neutron star core superfluid will pin against the quantized flux lines of the proton superconductor. As the neutron star spins down in the wind accretion phase of binary evolution, outward motion of vortex lines will reduce the dipole magnetic moment  in proportion to the rotation rate. The presence of a toroidal array of flux lines makes this mechanism inevitable and independent of the angle between the rotation and magnetic axes. The incompressibility of the flux-line array (Abrikosov lattice) determines the epoch when the mechanism will be effective throughout the neutron star. Flux vortex pinning will not be effective during the initial young radio pulsar phase. It will however be effective and reduce the dipole moment in proportion  with the rotation rate during the epoch of spindown by wind accretion as proposed by Srinivasan et al. The mechanism operates also in the presence of vortex creep. 
\end{abstract}

%%insert keywords separated by 3 hyphens using \keywords{words}

}]
%%close the twocolumn escape here

%%include \doinum{number}for the DOI number in the header
%%include \volnum{number} for the volume number in the header
%%include \year{yyyy} for  year of publication in the header
%%include \pgrange{num--num} page range of article in the header
%%include \artcitid{num} for the article citation id
%%include \lp to print last page of the article
%%include \setcounter{page}{pagenum} for the exact starting page of the article

\doinum{10.1007/s12036-017-9473-6}
\artcitid{44}
\volnum{38}
\year{September 2017}
%%\pgrange{23--25}
%%\setcounter{page}{23}
%%\lp{25}

\section{Introduction}
I first met Srini in 1973 when I joined the Theory of Condensed Matter group at the 
Cavendish Laboratory as a PhD student of Phil Anderson. Srini was a highlight of the company with his bright ideas and bright smile. After 1975 I saw him only a few times at conferences. Two of his seminal contributions have been of particular interest for me. 

The first was the paper with Radhakrishnan immediately after the discovery of the ﬁrst millisecond pulsar PSR B 1937+21 (Backer et al 1982). Immediately after the discovery two groups independently explained millisecond pulsars as the result of accretion in low mass X-ray binaries (LMXBs) (Alpar, Cheng, Ruderman \& Shaham 1982; Radhakrishnan \& Srinivasan 1982). Both groups made the bold inference that the LMXB magnetic fields must be as weak as $10^{8} {-} 10^{9}$ G in order to lead to millisecond equilibrium rotation rates, and predicted that the period derivative $\dot{P}$ of the millisecond radio pulsar would be as low as 10$^{{-}19}$ s s$^{{-}1}$. This prediction was soon verified by the measurement of $ \dot{P} = 1.2 \times 10^{{-}19} \rm{s\; s} ^{{-}1}$ from PSR B 1937+21 (Backer, Kulkarni \& Taylor 1983) indicating $10^8 {-} 10^9$ G fields in millisecond pulsars.

I reviewed (Alpar 2008) the arguments of these papers on the 10th anniversary of the eventual discovery of the first accreting millisecond X-ray pulsar (AMXP; Wijnands \& van der Klis 1998). In the Alpar et al paper we started from the expectation that neutron stars in LMXB have weak magnetic fields. Millisecond periods would be attained as the equilibrium periods with typical LMXB accretion rates $\dot{M} \sim 0.1 {\dot{M}}_{Edd} $ if the dipole magnetic field of the neutron star were $B \sim 10^9 $ G. At the end of accretion a millisecond pulsar would emerge with a period derivative $\dot{P} \sim 10^{-19}$ s s$^{-1} $, on the "spin-up" or "birth" line in the $P−\dot{P}$ diagram, shown for the first time in our paper.  Radhakrishnan \& Srinivasan (1982) started their argument by noting the lack of a supernova remnant, or any X-ray emission, from a nebula powered by the pulsar. If the millisecond pulsar had a conventional $10^{12} $ G magnetic field it would be very young, and would be associated with a supernova remnant and a pulsar wind nebula. Using the observational upper limits on the x-ray luminosity of the source they deduced  empirically that the dipole magnetic field must be less than about $4 \times 10^8$ G and $\dot{P} $ must be less than about 10$^{{-}19}$ s s$^{{-}1}$. They then noted that such a weak magnetic field would yield spin-up to a millisecond rotation period as the equilibrium period after accretion in a binary system. Our two groups independently arrived at the same picture tracing the available clues in different orders. 

After the discovery of the first accreting millisecond X-ray pulsar, the LMXB SAX 1808.4-3658, by Wijnands \& van der Klis (1998) others were discovered, including some that make transitions between X-ray and radio epochs (Papitto et al 2013). The connection between LMXB/AMXP and millisecond radio pulsars is now firmly established. Two basic questions arise regarding the evolution of these systems: (i) How come some millisecond pulsars are now no longer in binaries? (ii) How come the magnetic fields of the millisecond pulsars and LMXB/AMXP are so weak compared to the magnetic fields of young neutron stars? To answer (i), there are established evolutionary scenarios that explain the demise of the companion and emergence of a single radio pulsar after the LMXB phase. Regarding the second question one class of explanations for the low magnetic fields of millisecond pulsars invoke the burial of the magnetic field under accreted material during the LMXB phase. This is somewhat conjecturally dependent on the accretion history.  

My second favorite Srini contribution gives an elegant and convincing  answer to the question: Why are the dipole magnetic fields of neutron stars in LMXB $\sim$ 10$^{-3}$ times weaker than the fields in young radio pulsars? Rotation powered pulsars seem to retain initial dipole magnetic fields of $\sim$ 10$^{12}$ G  throughout their active pulsar lifetimes of $\sim $ 10$^6$ - 10$^7$ years. What happens to a neutron star in a binary during the subsequent 10$^{8}$ - 10$^{9}$ yr lifetime of evolution culminating in the LMXB phase to cause such a reduction of the dipole moment? Srini et al proposed that this field reduction is due to the pinning of quantized vortex lines in the neutron star core superfluid to the flux lines in the proton superconductor (Srinivasan, Bhattacharya, Muslimov \& Tsygan 1990 - SBMT). 

The neutron superfluid in the core of the neutron star takes part in the spindown by sustaining a flow of quantized vortex lines in the direction away from the rotation axis. The protons in the core are expected to be in the Type II superconducting phase (Baym, Pethick \& Pines 1969). Due to the pinning between vortex lines and the flux lines of the Type II proton superconductor in the neutron star core flux lines would be carried outward by the vortex lines. Thus flux would be expelled as the neutron star spins down under external torques. Spindown would induce a reduction in the dipole magnetic field in the same proportion as the reduction in the rotation rate: 
\begin{equation}
\frac{B(t)}{B(0)}=\frac{\Omega(t)}{\Omega(0)} .
\end{equation}
This depends on the core being entirely in a phase of superfluid neutrons coexisting with Type II superconducting protons. If parts of the core contain normal matter or Type I superconducting protons, these parts might relax their magnetic fields on shorter timescales, as the magnetic field in the crust regions is expected to do, and the long term evolution of the dipole surface field could still be governed by the SBMT mechanism, Eq. (1). In any case we will assume with SBMT, that Type II superconductivity and flux-vortex pinning are indeed the dominant features governing the magnetic field evolution of the neutron star. 

In the earlier epoch of binary evolution preceding the LMXB phase, the neutron star spins down while accreting from the stellar wind of its companion star to periods P $\sim $ 100 - 1000 s seen in high mass X-ray binaries with wind accretion, like Vela X-1.  Wind spindown from typical pulsar periods P $\sim $ 0.1 s to these periods would cause a reduction in the dipole magnetic field by a factor of $\sim $ 1000, yielding B $\sim $ 10$^{9}$ G at the start of the LMXB epoch. In Section 2 I will comment on flux line vortex line pinning and creep against this pinning, with emphasis on the effects of toroidally oriented flux lines. In Section 3 I will discuss the application of the scenario with spindown by wind accretion. 

\section{Flux - Vortex Pinning}
The possibility of flux line - vortex line pinning was briefly noted by Muslimov \& Tsygan (1985). Jim Sauls pointed out the importance of this for neutron star dynamics in his lecture in the 1988 NATO ASI on ``Timing Neutron Stars" (Sauls 1989). In his 1989 review on ``Pulsars: Their Origin and Evolution" Srini underlined the importance of this coupling not only for explaining field decay in the evolution leading to millisecond pulsars, but also for explaining why the field does not decay all the way to zero but has the typical value  B $\sim$ 10$^9$ G in the old population: ``A third and interesting possibility is implicit in the paper by Muslimov and Tsygan (1985), although not exploited by them. There are two sets of vortices in the quantum fluid interior: the magnetic vortices referred to above, and the vortices in the neutron superfluid. Muslimov and Tsygan suggest that the magnetic vortices could get pinned to the normal cores of the superfluid vortices." Srinivasan et al were the first to discuss in detail the evolutionary importance of this coupling in their seminal paper (1990; SBMT). 

For a poloidal orientation of flux lines there are easy directions along which the motion of neutron vortex lines away from the rotation axis required for the spin-down of the neutron superfluid will proceed without encountering pinning against flux lines. In other directions vortex motion will have to encounter flux lines and proceed via creep over the flux vortex pinning junctions by thermal activation. This problem of vortex creep over poloidal flux lines was addressed by Sidery \& Alpar (2009). As is generally true for magnetohydrodynamic stability, the flux line distribution in neutron stars is likely to include a toroidal component. Topologically unavoidable vortex pinning and creep against toroidal flux lines was addressed by G\"{u}gercino\u{g}lu \& Alpar (2014) and by  G\"{u}gercino\u{g}lu (2017). 

The magnetic dynamics of the Type II superconductor in coupling with the rotational dynamics of the neutron superfluid is a complicated problem for which the detailed solution on all different timescales is not known (Passamonti et al 2017). The essentials relevant for the evolutionary scenario of Srinivasan et al (1990) were presented in an important paper by Ruderman, Zhu \& Chen (1998). I will try to review and clarify the arguments of this paper, which develops a criterion for vortex - flux line pinning to effectively lead to flux decay induced by spindown, as proposed by SBMT. 

Ruderman, Zhu \& Chen start by noting that the other forces sustaining currents on a magnetic system should balance the Lorentz force on a macroscopic volume, yielding
\begin{equation}
\bf{F}=\frac{\bf{J} \times \bf{B} }{c}.
\end{equation}
 With $\bf{J} = \sigma \bf{E}$ and $\bf{E} = - ({\bf v} \times {\bf B }) / c $ one obtains the relation
\begin{equation}
v \cong \frac{F c^2}{\sigma B^2} 
\end{equation}
between the relative velocity $v$ of charges and magnetic flux, the force per unit volume $F$, the conductivity $\sigma$ and the magnetic field $B$. For the Type II superconductor they replace the magnetic field with macroscopic average $<B> = n_{\Phi} \Phi_0$, where $n_{\Phi}$ is the area density of flux lines and $\Phi_0 = hc/2e$ is the flux quantum, to obtain
\begin{equation}
v_{\Phi} \cong \frac{F c^2}{\sigma_{eff} (n_{\Phi} \Phi_0)^2} 
\end{equation}
for the relative speed of the flux lines with respect to the charges (the center of mass of the electron - superconducting proton plasma), in terms of the effective conductivity $\sigma_{eff}$. The drag force due to the scattering of electrons from the magnetic field in the cores of the flux lines drives the decay of the magnetic flux by the flow of the flux lines at speed $v_{\Phi}$ away from the magnetic axis (or the circular axis of the toroidal distribution in the case of toroidal flux lines). This force (per unit length of flux line) is a linear drag force
\begin{equation}
\bf{f} = - \eta \; (\bf{v}_c - \bf{v}_L)
\end{equation}
where $\bf{v}_c $ and  $\bf{v}_L$ are the velocities of the electrons and the flux line respectively. The coefficient $\eta$ has been calculated for scattering of electrons from a magnetized neutron vortex line (Alpar, Langer \& Sauls 1984), and can readily be scaled for the scattering of electrons from a flux line. This drag force will govern the motion of the flux line with respect to the proton superfluid/superconductor, of velocity ${\bf{v}_s}^p $, through the Magnus equation, or equivalently, the flux line moves with respect to the background proton superfluid such that the drag force balances the Lorentz force on the flux line: 
\begin{equation}
{\bf{f}} = \frac{n_{p}\; c}{e}\; \bf{\Phi}_0 \times ({{\bf{v}}_s}^p  - {\bf{v}}_L) = \rho_p {\bf{\kappa}} \times ({{\bf{v}}_s}^p - {\bf{v}}_L) .
\end{equation}
Here $\rho_p $ and $n_p = n_e$ are the mass and number densities of the superconducting protons, 
${\bf {\kappa}}$ and $\bf{\Phi}_0 $ are vectors directed along the flux line, with magnitudes 
$\kappa = h/2m_p$  and $\Phi_0 = hc/2e$, the quanta of vorticity and flux, respectively. 
From Eqs.(5) and (6) one obtains the flux line velocity $v_{\Phi, r}$ in the direction 
away from the magnetic axis
\begin{equation}
v_{\Phi, r} =  \alpha ({v_s}^p - v_c)
\end{equation}
where $\alpha$ is given by the ratio and inverse ratio of the inertial and drag coefficients:
\begin{equation}
\alpha = \left[\frac{\rho_p \kappa}{\eta} + \frac{\eta}{\rho_p \kappa}\right]^{-1} = \left[\frac{n_e e \Phi_0}{\eta c} + \frac{\eta c}{n_e e \Phi_0}\right]^{-1} .
\end{equation}
The effective conductivity $\sigma_{eff} $ of Ruderman, Zhu \& Chen (1998) can now be 
obtained in a quick and illuminating manner. The rotational (solenoidal) electric 
field ${\bf E} = -1/c (\partial{\bf A}/\partial{t})$ which plays the leading role in the dynamics of magnetic field decay (Passamonti et al 2017) is set up by the radial flow of flux lines in the case of a Type II superconductor:
\begin{equation}
E =   - \frac{1}{c} \frac{\partial{A}}{\partial{t}} = \frac{n_{\Phi}\Phi_0}{c}v_{\Phi, r}.
\end{equation} 
Using Eq.(7), 
\begin{equation}
E =   \frac{\alpha n_{\Phi}\Phi_0}{n_e e c} j \equiv \frac{j}{\sigma_{eff}}
\end{equation} 
we obtain 
\begin{equation}
\sigma_{eff} = \frac{n_e e c}{\alpha n_{\Phi}\Phi_0} = \frac{n_e e c}{n_{\Phi}\Phi_0} \left[\frac{n_e e \Phi_0}{\eta c} + \frac{\eta c}{n_e e \Phi_0}\right] .
\end{equation} 

The expression for the drag coefficient $\eta$ on flux lines is scaled from the 
corresponding expression for the drag coefficient on magnetized vortex 
lines (Alpar, Langer \& Sauls 1984). The magnetized vortex line has a flux $\Phi_{*} $ 
and a London radius $\Lambda_{*} $ which depend on the superfluid proton flow 
dragged around the neutron vortex line and therefore on the effective neutron and proton masses. 
For electron scattering from the flux line these quantities are $\Phi_0 $ and the 
London radius $\Lambda$ for the proton superfluid, giving
\begin{equation}
\eta \cong 1.3 \times 10^{-2}\; (\rho_{p,14})^{1/6} \rho_p \kappa  
\end{equation}
where $\rho_{p,14}$ is the superconducting proton density in units of 10$^{14}$ g cm$^{-3}$. With $\rho_p \kappa \gg \eta$ we  obtain
\begin{equation}
\sigma_{eff} = \frac{(n_e e)^2}{\eta n_{\Phi}}
\end{equation}.

Now we turn to the flux line - vortex line pinning. Flux lines (and vortex lines pinned to them) 
are moving with the velocity $\bf{v}_c + \bf{v}_{L, r}$. To remain pinned, 
the neutron vortex lines would have to be moving at the velocity 
$\bf{v}_c + \bf{v}_{L, r} - {\bf{v}_s}^n $ with respect to the background 
neutron superfluid. According to the Magnus equation of motion such a nonzero 
velocity difference between vortex lines and background neutron superfluid is 
sustained by the pinning forces arising at flux - vortex junctions. 
Pinning cannot be sustained if the velocity difference exceeds the 
maximum (critical) velocity difference corresponding to the maximum pinning force available 
at the junction. The direction of $\bf{v}_{\Phi, r}$ is radially outward from the magnetic dipole axis, 
or outward from the circular symmetry axis of the toroidal distribution of flux lines, as the case 
may be. $\bf{v}_c$ and ${\bf{v}_s}^n $ are in the direction of the rotational flow, i.e. 
the azimuthal direction with respect to the rotation axis. Leaving directional aspects aside, 
pinning can be sustained, and so flux lines will move together with vortex lines in directions 
radially outward from the rotation axis as the star spins down, 
if ${v}_{\Phi, r} \lesssim v_{cr}$. The critical velocity $v_{cr}$ for a 
vortex line to remain pinned to a flux line, corresponding to the 
maximum pinning force per unit volume $F_{max}$ is given by Eq. (4 ):
\begin{equation}
v_{cr} \cong \frac{F_{max} c^2}{\sigma_{eff} (n_{\Phi} \Phi_0)^2} .
\end{equation}

The pinning energy is determined by the magnetic energies in the cores of the 
flux lines and the spontaneously magnetized vortex lines. The magnetic cores 
of the flux and vortex lines carry fluxes $\Phi_0$ and $\Phi_*$ and have 
radii $\cong \Lambda$ and  $\Lambda_{*}$ respectively. In the following 
estimations we will neglect factors $\sim $ O(1) that depend on effective nucleon masses, 
and adopt $\Phi_{*} = \Phi_0$ and $\Lambda_{*} = \Lambda$, the London penetration depth 
of the proton superconductor. 
For both superfluids $ \Lambda > \xi_i$, the coherence length of the neutron or proton superfluid, 
and the magnetic energy in the flux line and magnetized neutron vortex line cores is larger 
than the condensation energy. 

The number of flux line-vortex line junctions per unit volume is 
\begin{equation}
n_{junc} = \frac{{(\pi \Lambda^2)}^2}{{l_{\Phi}}^2 {l_{vort}}^2 V_{junc}} 
\end{equation}
where $l_{\Phi}$ and $l_{vort}$ are the average spacings between flux and vortex lines respectively, 
and $V_{junc} \sim \Lambda^3 $ is the volume of a flux-vortex junction; 
${l_{\Phi}}^2 = \Phi_0 / B $ and ${l_{vort}}^2 = 2\Omega / \kappa$. 
Note that $n_{junc}$ has the same dependence 
on the flux line and vortex line densities.   
 
The pinning energy at each flux-vortex junction is estimated as
\begin{equation}
E_{pin} \sim \frac{1}{8\pi} (\frac{\Phi_0}{\pi \Lambda^2})^2 V_{junc} ln(\Lambda/\xi)
\end{equation}
The pinning energy is the magnetic energy gain in the overlap 
volume at the vortex line-flux line junction. 
The factor $ln(\Lambda/\xi) \cong 5$ 
accounts for the local current structure around the flux 
and vortex lines. The maximum pinning force per junction 
is given by $F_{pin} = E_{pin}/\Lambda $ 
where $E_{pin}$ is the pinning energy at each flux - vortex junction. 
This leads to the maximum force per unit volume
\begin{equation}
F_{max} = F_{pin}\; n_{junc}  \sim \frac{1}{8\pi}\; \frac{{\Phi_0}^2} 
{{l_{\Phi}}^2 {l_{vort}}^2 }\; \frac{1}{\Lambda}\;ln(\Lambda/\xi).
\end{equation}
The result of Ruderman, Zhu \& Chen (1998) is larger than 
this by a factor ${l_{\Phi}}^2 /\Lambda^2  $ because they 
take the spacing between the flux lines swept by a 
vortex line to which they are pinned to be $\sim \Lambda$, the minimum spacing allowed before 
the superconducting phase is lost at the higher critical field. Let me try to clarify this very high 
density of flux lines around the vortex line. Ruderman, Zhu \& Chen write in the Appendix to 
their paper that
``the typical distance between two
consecutive flux tubes pushed by the same moving vortex is 
about $\Lambda$. ... The magnetic repulsion between flux 
tubes limits their density. This repulsion is not effective 
until the inter-flux tube
separation approaches $\Lambda$." This enhanced density of flux lines carried 
along by a vortex line must be very local to the vicinity of the vortex line, 
which is, so to speak dressed by the cluster of pinned flux lines it has picked up 
during its entire journey through the star. The background configuration of the 
flux line array cannot have been effected. The 
extra energy cost of a non-uniform macroscopic average B field 
would be prohibitively large if the pile up of flux lines plowed 
along by a vortex line extended to distances greater than $l_{\Phi }$. 
Thus as a vortex line moves the extra flux line density it carries along must be confined to 
distances of order $\Lambda$ in directions transverse to the vortex line. 
Other than this the flux line array between two vortex lines will be 
pushed forward by the vortex on one side and pulled along 
by the vortex on the front while retaining the equilibrium density $n_{\Phi}$, i.e. without being compressed. 
To check the consistency of this picture, we note that the vortex line needs to pick up 
$l_{\Phi }/\Lambda$ flux lines from each layer of the flux line array it encounters, 
corresponding to a fraction ${l_{\Phi }}^2/R \Lambda \sim 2 \times$ 10$^{-14}$, taking the 
neutron star radius R $\sim $ 10$^6$, B$\sim $10$^{12}$ G and $\Lambda \sim $100 fm. 
This small fraction of flux lines are those that are dislodged from the strong binding of 
the Abrikosov lattice, corresponding to the strong pinning regime described by 
Alpar, Anderson, Pines \& Shaham (1984) in the context of vortex pinning in the neutron star crust lattice.  

These considerations justify the estimate of Ruderman, Zhu \& Chen (1998), 
\begin{equation}
F_{max} \cong \frac{1}{8\pi}\; \frac{{\Phi_0}^2} 
{{l_{vort}}^2}\; \frac{1}{{\Lambda}^3}\; ln(\frac{\Lambda}{\xi})
\end{equation} 
From Eqs. (14) and (18) one obtains 
\begin{equation}
v_{ cr} = 2 \times 10^{-10} \; \frac{\Omega}{B_{12}} \; {\rho_{p,13}}^{2/3}\;ln(\frac{\Lambda}{\xi}) \; {\rm cm / s} 
\end{equation}
where $\rho_{p,13}$ is the density of superconducting 
protons in units of $10^{12} g cm^{-3}$. 

For the steady state spin-down of the neutron superfluid 
at the rate $\dot{\Omega}$ dictated by the external torque, 
the average radial velocity of vortex lines at distance $r$ 
from the rotation axis is given by 
\begin{equation}
v_{vort, r} = \frac{|\dot{\Omega}| r}{2\Omega}
\end{equation}
This average macroscopic velocity is due to all microscopic dissipative interactions between the normal matter core of the vortex line and ambient normal matter like e.g. electrons scattering from the vortex core. Flux lines and vortex lines will remain pinned and move together 
as long as $v_{vort, r }$ remains less than the critical 
velocity, $v_{vort, r} (r) <  v_{cr}$. This will hold at 
distances $r$ from the rotation axis satisfying
\begin{equation}
r < r_{cr} \cong 4 \times 10^{-10}  \frac{\Omega^2}
{|\dot{\Omega}| \; {B_{12}} }\; {\rho_{p,13}}^{2/3}\; ln(\frac{\Lambda}{\xi})\; {\rm cm}.  
\end{equation}
For pulsars older than the Vela pulsar ( age greater than 10$^4 $ years) $r_{cr}\gtrsim $ 10$^6 $ cm, so that flux is being gradually expelled from the core of the neutron star. However, flux expelled from the core is not able to diffuse through the neutron star crust during the pulsar phase, as the magnetic field diffusion timescales through the high conductivity crust are estimated to be $\sim $10$^7 yr$. Indeed the observed dipole moment distribution of young radio pulsars shows no indication of field decay. 

At finite temperature vortex and flux lines will creep across pinning energy 
barriers by thermal activation. This process will allow the outward flow of 
vortex lines at the velocity $v_{vort,r}$ to achieve its steady state value, given in  Eq.(20), 
defined by the external torque. This motion of the vortex lines in turn allows the superfluid 
to spin down at the steady state rate $\dot{\Omega}$. The velocity difference $v_{\infty} $ 
between $\bf{v}_c$ and ${\bf{v}_s}^n$ in steady state creep is always less than $v_{cr}$ 
so vortex creep will always operate when conditions allow pinning. Recent work interpreting 
certain components of postglitch relaxation in the Vela pulsar and many older pulsars in terms 
of vortex creep against toroidal flux lines 
(G\"{u}gercino\u{g}lu \& Alpar 2014; G\"{u}gercino\u{g}lu 2017) supports the conclusion that 
flux-vortex pinning and therefore SBMT flux decay induced by spin down do operate in these pulsars.

\section{Field Decay under Wind Accretion}

Srinivasan et al (1990) reasoned that the long term spindown of the neutron star by the companion's wind during the detached epoch of binary evolution is the decisive stage of evolution for the reduction of the average dipole magnetic field from B $\sim$ 10$^{12}$ G to B $\sim$ 10$^{9}$ G as the neutron star is spun down from $\Omega_1 \sim$ 1 rad s$^{-1}$ when wind accretion starts at the end of the pulsar/ejector phase, to $\Omega_2 \sim$ 10$^{-2} - $10$^{-3}$ rad s$^{-1}$ as exemplified by systems like Vela X-1. Patruno et al (2012) applied the SBMT scenario  to the spin and magnetic field evolution of the accreting X-ray pulsar IGR J17480-2446 in Terzan 5. The neutron star was evolved through the wind accretion phase assuming B(t) $\propto \Omega(t)$ due to flux-vortex pinning. 

After the pulsar crosses the death valley and pulsar activity stops the neutron star continues to spin down under the dipole spindown torque until wind accretion starts when the Alfven radius reaches the light cylinder. This happens at a rotation rate 
\begin{equation}
\Omega_1 =  6.4\; {\mu_{29}}^{-4/7}\; {\dot{M}_{11}}^{2/7}\;{(\frac{M}{1.4 M_{\odot}})}^{1/7} {\rm rad \; s}^{-1}.
\end{equation}
where $\mu_{29}$ is the initial dipole moment of the neutron star, preserved through the initial radio pulsar phase, in units of 10$^{29}$ G cm$^{3}$ and $\dot{M}_{11}$ is the rate of mass capture by the neutron star from the wind in units of $10^{11 } g s^{-1} $. The neutron star mass $ M$  is given in units of 1.4 solar masses $M_{\odot} $. The wind mass loss rate of a solar mass main sequence star is about 10$^{12}$ g s$^{-1} $ and a few percent of this is expected to be captured by the neutron star (Nagae et al. 2001; Theuns et al 1996), indicating mass accretion rates in the 10$^{10 }$ - 10$^{11}$ g s$^{-1} $  range.
The X-ray flux distribution of galactic neutron stars binaries undergoing wind accretion (Pfahl, Rappaport \& Podsiadlowski 2002) also indicates accretion rates in the $10^{11 } g s^{-1} $ range.  (The wind accretion rate employed for  IGR J17480-2446 by Patruno et al (2012) is two orders of magnitude larger in view of the large rotation rate expected for the synchronously rotating companion  in that relatively young system).  The LIGO discovery of gravitational radiation from the merging of two $\sim$ 30 $M_{\odot}$  blackholes ((Abbott et al 2016a) implies weak winds for massive progenitor systems, especially in old, low metallicity populations (Abbott et al 2016b). If  weak winds are also common among few solar mass main sequence companions  especially with low metallicity, binaries with  $\dot{M} \sim$ 10$^{10 }$ 
g s$^{-1} $ may be common in old population environments which host the progenitors of LMXB and millisecond pulsars.

The wind accretion onto the neutron star produces a spin-down rate
\begin{equation}
\dot{\Omega} = 4.9 \times 10^{-17}\;{ I_{45}}^{-1}\; 
\dot{M}_{11}\;{\mu_{29}}^{2/7}\; {(\frac{M}{1.4 M_{\odot}})}^{3/7} {\rm rad\; s}^{-2}.
\end{equation}
where $ I_{45} $ is the moment of inertia of the neutron star in units of 10$^{45}$ g  cm$^{2}$.  
Substituting these values in Eq. (21) one can check for consistency: 
\begin{eqnarray}
r_{cr} \cong 6.6 \times 10^{7} \; {\mu_{29}}^{-10/7}\; {\dot{M}_{11} }^{-3/7}\; {(\frac{M}{1.4 M_{\odot}})}^{-1/7}\;{ I_{45}} \\ \nonumber
\times (\rho_{p,12})^{2/3}\;  ln(\frac{\Lambda}{\xi})\;{\rm cm}.
\end{eqnarray}
Thus throughout the wind accretion era $r_{cr}$ is larger than the neutron star radius, so that flux vortex pinning prevails and spin-down induces field decay. 

I conclude that the neutron star dipole magnetic 
moment will be reduced through the SBMT mechanism, as the star spins down by 
wind accretion.  Solving Eq. (23) with Eq.(1) leads to 
the spindown and field decay timescale
\begin{eqnarray}
t_{sd} & = & \frac{7}{5}\; \frac{I \Omega_1}{ {\mu}^{2/7}\; {\dot{M} }^{6/7}\; {(GM)}^{3/7}}  \nonumber \\
& \cong & 6 \times 10^9 \; {\mu_{29}}^{-6/7}\; {\dot{M}_{11} }^{-4/7}\;{(\frac{M}{1.4 M_{\odot}})}^{-2/7} {\rm yr} .
\end{eqnarray}

\section{Conclusion}
The very simple and elegant mechanism proposed by Srinivasan et al (1990) explains the comparatively weak magnetic dipole moments of millisecond pulsars by spin-down induced flux decay due to flux line - vortex line pinning. The mechanism will work during the long wind accretion phase of binary evolution. This is the crucial evolutionary phase for explaining the 1000-fold reduction in magnetic moments from the young radio pulsars to the old populations of LMXB, accreting X-ray millisecond pulsars and millisecond radio pulsars. Thus the SBMT mechanism indeed produces the weak fields needed for the final LMXB spin-up to millisecond periods. I wish Srini a very happy 75th birthday and congratulate and thank him for his many brilliant contributions.

\section*{Acknowledgments}
I thank Erbil G\"{u}gercino\u{g}lu for his careful reading and suggestions and Onur Akbal for help with the manuscript. The author is a member of the Science Academy, Turkey.

\end{document}